# Electro-capillary peeling of thin films

Peiliu Li[1,2], Xianfu Huang[1,2] & Ya-Pu Zhao[1,2 ✉]

Thin films being a universal functional material have attracted much interest in academic and industrial applications, such as flexible electronics, soft robotics, and micro-nano devices. With thin films becoming micro/nanoscale, developing a simple and nondestructive peeling method for transferring and reusing remains a big challenge. Here, we present an innovative detaching approach for thin films: Electro-capillary peeling method. The electro-capillary peeling method achieves thin films' detachment by driving liquid to percolate and spread into the bonding layer under electric fields. Compared with traditional methods, thin film detached by this novel peeling mode shows a much lower deformation and strain of film (reaching 86%). Evaluated by various applied voltages and films, the electro-capillary peeling method shows active control characterizations and is appropriate in a broad range of films. Theoretically, it is demonstrated that the electro-capillary peeling method is actualized by utilizing Maxwell stress to compete with the film's adhesive stress and tension stress. The peeling length versus time, applied voltage, film's thickness, and elastic modulus are described by $r \sim t$, $r \sim U$, $r \sim d_0^{-1/2}$, and $r \sim E^{-1/2}$, respectively. Additionally, observations of critical peeling voltage present that the thin film is readily detached by using the electro-capillary peeling method at the ultra-low voltage of only 0.7 V. This work clearly shows the great potential of electro-capillary peeling method to provide a new way for transferring films and open novel routes for applications of soft materials.

[1] State Key Laboratory of Nonlinear Mechanics, Institute of Mechanics, Chinese Academy of Sciences, Beijing, China. [2] School of Engineering Science, University of Chinese Academy of Sciences, Beijing, China. ✉email: yzhao@imech.ac.cn

Thin films being a universal functional base material with a variety of outstanding properties have been widely investigated and used in fundamental studies and applications, including flexible electronics[1, 2], soft robotics[3, 4], micro-/nano-electro-mechanical systems (MEMS/NEMS)[5, 6], and biomedical devices[7, 8]. Planted by designed structures and devices, thin films are achieved to a specific function. For example, carbon nanotube thin-film transistor is a promising candidate for flexible and wearable electronics[9, 10], thin films located with a crystalline can be programmed into various soft robotics[11, 12], and thin film is planted with ZnO nano-rods to fabricate self-cleaning surface[13, 14]. Because thin films and planted structures/devices often have greatly different mechanical properties, such multilayered thin film systems are often prone to interfacial delamination/crack under various loading conditions resulting in premature failure of the devices. Therefore, investigating a nondestructive peeling method for thin films is quietly attractive for the long-term reliability of functional devices in applications.

With the rapid development of miniaturized components, the thickness of film becomes micro/nanoscale. More recently, the capability of micro/nanofilms presents an exciting opportunity in materials design, such as assembling multiple two-dimensional (2D) materials with complementary properties into layered heterogeneous structures[15, 16], superconducting films for electronic devices[17], and supercapacitor electrodes in energy storage[18, 19]. Yet, the traditional peeling methods, including debonded strip[20] and blister[21], are difficult to satisfy the needs of practical productions at such scales, especially for the fully-attached micro/nanofilm. For the sake of practical applications, several strategies have been pursued to manipulate the conditions for detaching the micro/nanofilm from the substrate, including hydrogen bubbles evolved at alkaline water electrolysis for graphene on the metallic catalyst[22], drying-induced peeling for colloidal films[23], water-soluble chemical bonding layer[24], and the reduction of surface adhesion modified by chemical/physical approaches[25-28]. Generally, these peeling strategies take advantage of modifying the bonding layer's properties between the film and substrate. Despite extensive progress, the development of peeling method for micro/nanofilms remains in infancy, owing to the complicated preparations and the limitation of applied films. Therefore, it is urgent to develop a simple peeling method for various thin films.

In contrast to complicated modifications of the bonding layer's properties, a thin liquid layer percolating and spreading into the bonding layer can serve as a simple physical peeling method to detach the film from the substrate. Capillary peeling approach is a passive method by utilizing the liquid layer to detach an attached film when the film-substrate interface contacts water[29, 30]. The principle of capillary peeling method illustrates that it is generally suitable for films with low surface adhesion, especially hydrophobic films. Practically, there are many films with a high adhesive force on the substrate, and the peeling rate in an on-demand manner is highly preferred, however, it poses stricter requirements on manipulation design. To this end, developing an active peeling method that enables easy detachment of various micro/nanofilms still remains a big challenge. Previous studies[31-33] have proven that the wettability of a liquid can be controlled by electric fields, including surface tension, contact angle, and movement of the contact line. Electric charge in the liquid is loaded by the Maxwell stress, which would impact the motion of liquid[34-36]. Both the wettability and motion of the liquid would be controlled by the electric field. Therefore, developing a peeling method by controlling a thin liquid layer to detach micro/nanofilms under an electric field is perhaps an innovative option.

Herein, we develop an electro-capillary peeling method for the thin film's detachment that is achieved by driving liquid to percolate and spread into the bonding layer under electric fields. Unlike the complicated modification of bonding layer and applied film restriction of capillary peeling, the electro-capillary peeling method is an active detaching approach with a simple apparatus and is appropriate in a broad range of films, even in hydrogels of high adhesive stress. Theoretically, it is demonstrated that the electro-capillary peeling method is actualized by the Maxwell stress competing with the film's adhesive stress and tension stress. The peeling length versus time, applied voltage, film's thickness, and elastic modulus are described systematically by $r \sim t$, $r \sim U$, $r \sim d_0^{-1/2}$, and $r \sim E^{-1/2}$, respectively. In addition, observations of critical peeling voltage indicate that thin films are easily detached by using the electro-capillary peeling method with an ultra-low voltage of only 0.7 V, and the thin film's deformation declares that the electro-capillary peeling method is a nondestructive approach for the functional device's protection. For the advantages, the electro-capillary peeling method is of significance for the promising application of thin film's detaching and transferring.

**Results**

**Experimental apparatus and characterization of electro-capillary peeling.**

Figure 1a shows a schematic drawing of the electro-capillary peeling method that mainly includes three core elements: a film, an electrolyte droplet, and a supply power. Initially, polydimethylsiloxane (PDMS) film served as tested samples for its stable property and extensive application, and it of a micron thickness was fabricated by using the spin coating method (see Methods). After being treated by the plasma, PDMS films were bonded on the indium tin oxide (ITO) glass surface with a firm and uniform bonding layer (Supplementary Fig. S1). The wettability illustrates that the bonded surface is hydrophilic, and the peeling test presents that the adhesive stress of bonding layer is about 7 N/m at the peeling rate of 1 mm/s. So clearly, there is a potent combination between PDMS films and ITO glass surfaces. Furthermore, a circular hole with a radius of 2 mm was cropped out at the center of bonded PDMS film. A micro pump injected an electrolyte droplet with a volume of 20 μL into this circular hole, and then an electric field was applied to the liquid droplet by a direct current (DC) supply power with two platinum (Pt) wire electrodes. One of the Pt wire electrodes was inserted into the liquid droplet and connected to the positive pole, and the other Pt wire electrode was pasted on the ITO glass surface and attached to the negative pole. After turning the power on, it is excitingly observed that the liquid droplet percolates into the bonding layer and detaches the PDMS film from the ITO glass (Fig. 1b and Supplementary Movie 1). Viewed from side views, the peel-off film's profile is similar to parabolic (Fig. 1c). Observed from top views, the wetting shape is an axisymmetric circle, and the ring's radius $r(t)$ is the peeling length, which increases with time (Fig. 1d). Under electric fields, the liquid is driven to percolate and spread into the bonding layer for films' detachment.

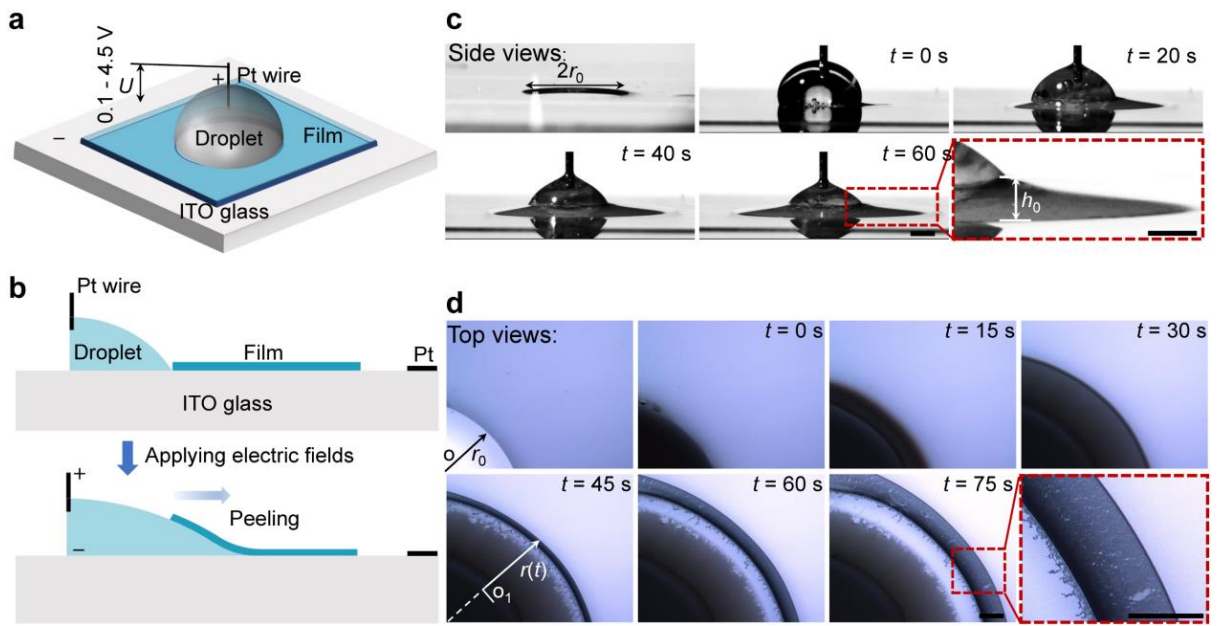

**Fig. 1| Experimental setup and characterization of electro-capillary peeling. a** Experimental apparatus setup. The electric field is applied through two Pt wire electrodes, one of the Pt wire electrodes inserted in the liquid droplet and the other placed on the ITO glass. **b** Schematic illustration of the working process for the electro-capillary peeling method. The liquid droplet is driven to percolate into the bonding layer and peel off the PDMS film from the ITO glass by applying a DC electric field. **c-d** Side and top views of electro-capillary peeling process. $r_0$ is the radius of the pre-fabricated hole on the PDMS film, $h_0$ is the lifting height at the edge of the film, and $r(t)$ is the time-dependent peeling length. Here, all scale bars are 1 mm.

**Characterizations of electro-capillary peeling at different voltages.**

To explore the impact of electric intensity on the electro-capillary peeling method, observation of lifting height and peeling length was conducted with broad ranges of applied voltages. As shown in Fig. 2a, side views of electro-capillary peeling present that the lifting height of PDMS film is similar at the voltage from 1.5 to 4.5 V. Yet, the droplet outline is quite different at various voltages. With the increase of applied voltage, the topography of droplet changes from a semicircle shape to a half ellipse due to the increased peeling length (Supplementary Movie 2). Figures 2b and c present top views of electro-capillary peeling in axisymmetric and planar peeling modes. These two modes are the main types in practical applications[37], and their experimental setups are shown in Supplementary Fig. S2. At 1.5 V, the peeling length is really small and even smaller than the radius of the pre-fabricated hole, although after 60 s. As the applied voltage increases to 4.5 V, the peeling length is almost three times as much as that at 1.5 V. In these two modes, observation of peeling length shows that both the efficiency and manner of them are similar (Supplementary Movie 3), and the peeling rate increases with increasing applied voltage. Statistical results of the lifting height and the peeling length at different voltages are presented in Figs. 2d, e. When the applied voltage changes from 1.5 to 2.5 V, the lifting height changes 122% (from 0.23 to 0.51 mm). As the applied voltage increases above 2.5 V, the lifting height varies inconspicuously with increasing voltage (Fig. 2d). Figure 2e shows the peeling length versus time at diverse voltages. At the same peeling time, the peeling length increases obviously with increasing applied voltage. The peeling length-time curve further presents that the peeling length is nearly linear with time. At the voltages of 1.5, 2.5, 3.5, and 4.5 V, the average peeling rate is 0.013, 0.023, 0.038, and 0.048 mm/s, respectively. The peeling rate shows a linear relationship with the applied voltage. Therefore, depending on the practical application, the efficiency of electro-capillary peeling would be actively controlled by the applied electric field.

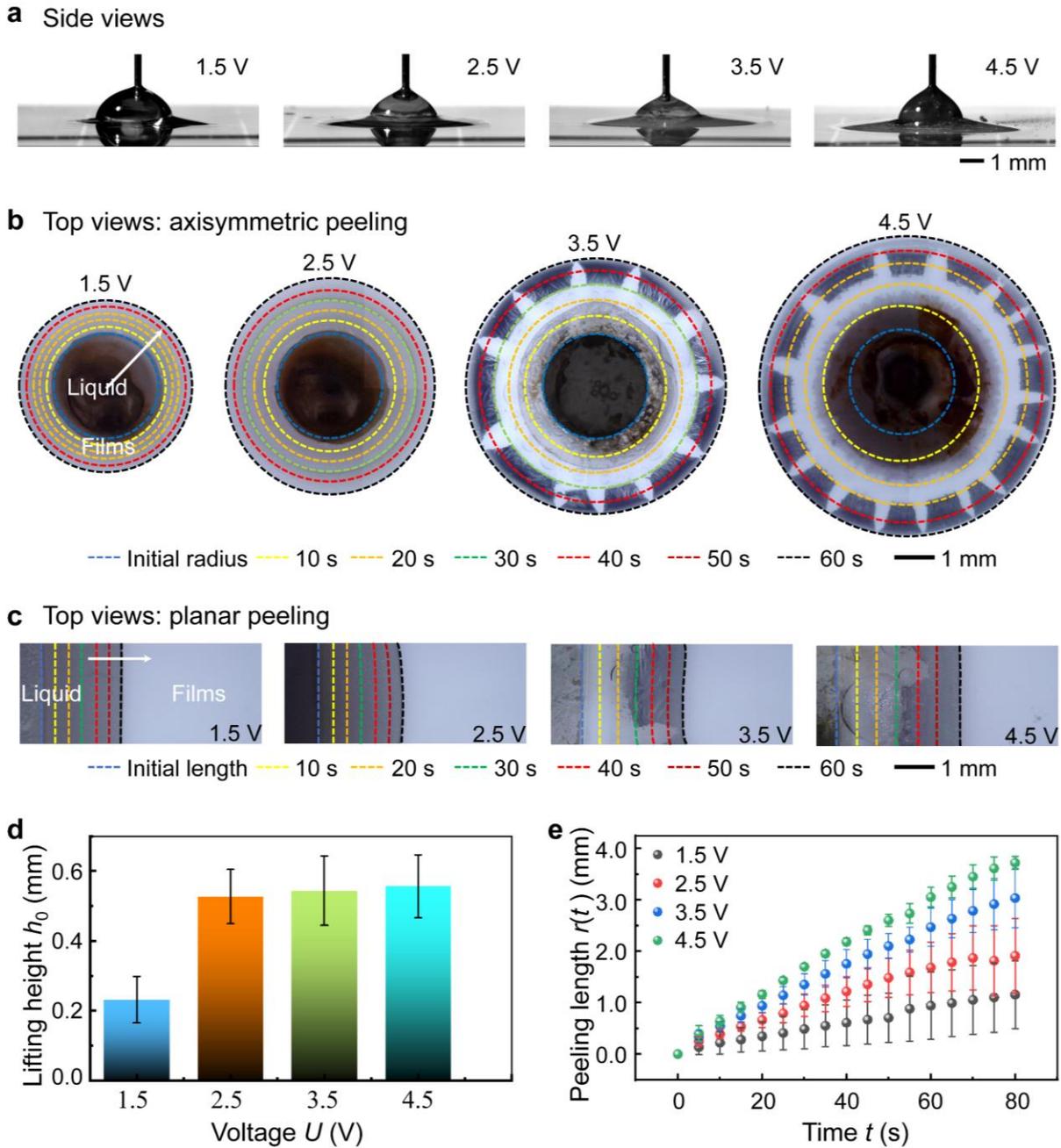

**Fig. 2| Electro-capillary peeling at different voltages. a** Side views of electro-capillary peeling at various voltages. **b-c** Top views of axisymmetric and planar peeling at diverse voltages. The selected photo is the presentation of the end of electro-capillary peeling ($t$ = 60 s), and dotted lines with different colors are used to mark its evolution. The zigzag graphics are the anode pattern formed by the electrode corrosion at 3.5 and 4.5 V. **d-e** Statistical results of lifting height and peeling length in the electro-capillary peeling method. At 1.5, 2.5, 3.5,

and 4.5 V, the lifting height is 0.23, 0.51, 0.52, and 0.54 mm, and the peeling length is 1.0, 1.8, 3.0, and 3.8 mm ($t$ = 80 s), respectively. Error bars in **d-e** are the standard deviation of raw data.

**Electro-capillary peeling method for various films' detachment.**

Practically, there are many films with various properties employed in applications. To evaluate the practicality of electro-capillary peeling method, systematic experiments were conducted with films of different thicknesses, elastic moduli, and types. Initially, PDMS films with thicknesses from 25 to 300 μm were fabricated by a designed spin coating rate (see Methods). A representative SEM image of PDMS film on the glass is shown in Fig. 3a, and PDMS films with thicknesses from 25 to 300 μm are presented in Supplementary Fig. S3. Fabricated using this way, PDMS films are of a uniform thickness and well bonded on the glass. Furthermore, PDMS films with the elastic modulus of 1.0, 1.6, and 2.4 MPa were achieved by consisting of the intentional weight ratio of base and curing agent (see Methods). Their force-displacement curves indicate that the film is at the elastic stage within the tensile range of 25 mm (Fig. 3b). Peeling test shown in Supplementary Fig. 4 reveals that the adhesive stress of these films on the glass is virtually identical. Ultimately, the functional film, including hydrogel, polyethylene terephthalate (PET), and polyethylene naphthalate (PEN) film, is also tested by using the electro-capillary peeling method. SEM images present that these functional materials' bonding layer is the same as that of PDMS films (Fig. 3c). Statistical results of lifting height and peeling length for various films are provided in Figs. 3d and e. Both the lifting height and peeling length decrease with increasing the film's thickness. When the film's thickness changes from 25 to 300 μm, the lifting height and the peeling length are reduced from 0.62 to 0.13 mm and from 3.61 to 0.97 mm, respectively. Observations of electro-capillary peeling with films of various elastic moduli are similar to that of thickness. The lifting height and peeling length decrease 89.3% and 64.0% when the elastic modulus increases from 1.0 to 2.4 MPa. For a thicker or harder film, the electro-capillary peeling rate would be slower and perhaps reduced to be ignored (Supplementary Fig. S5). For widely used functional films (hydrogel, PET, and PEN film), the electro-capillary peeling method also detaches them from the substrate, although they have variegated elastic moduli and adhesive stress (Supplementary Table 1). Observations present that the lifting height of hydrogel is the largest while that of PET is the smallest, and the peeling length of hydrogel is the smallest but that of PEN is the largest. Although the performance of electro-capillary peeling method is affected by the film's properties, this approach is clearly appropriate in a broad range of films.

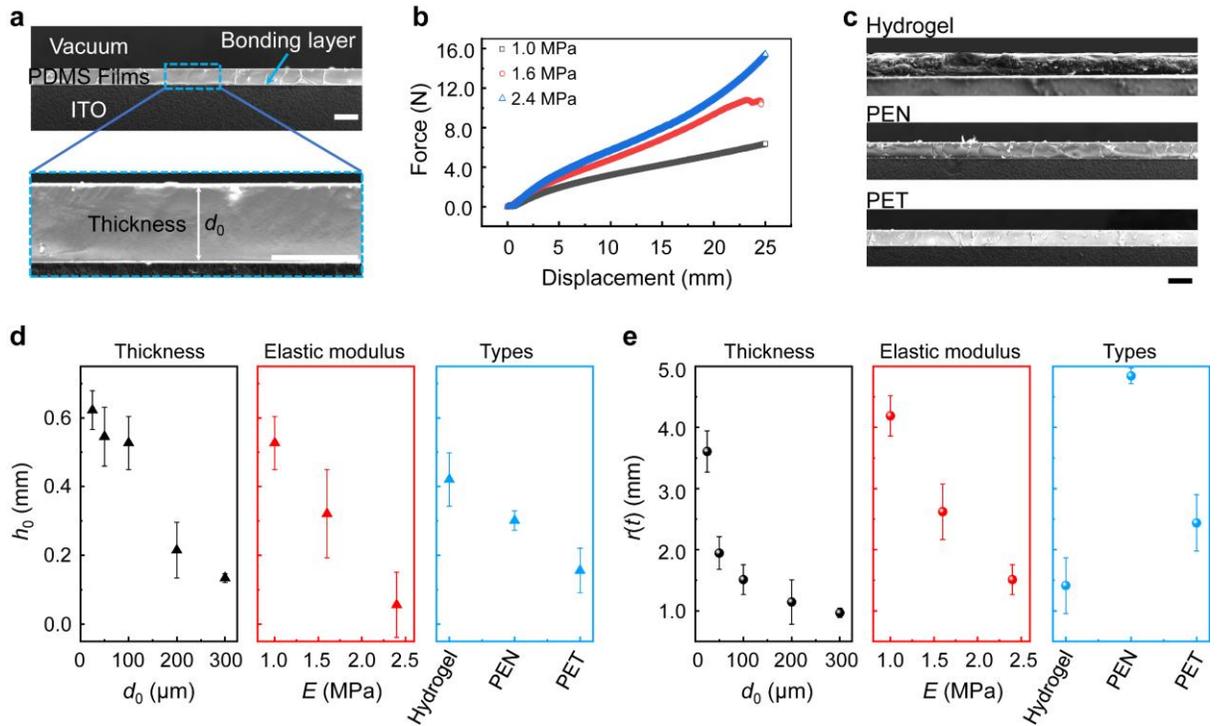

**Fig. 3| Electro-capillary peeling for various films. a** Representative SEM image of PDMS film on the glass surface. $d_0$ is the film's thickness. **b** Force-displacement curves of the film with different elastic moduli. The thickness and width of the tested film are 1 mm and 20 mm, respectively. **c** SEM images of functional films on the glass surface. The thickness of Hydrogel, PEN, and PET film are all 100 μm. **d-e** Statistical results of lifting

height and peeling length for films with diverse thicknesses, elastic moduli, and types. These statistic results are acquired at 80 s during the peeling process. The error bars in **d-e** are the standard deviation of raw data, and all scale bars in **a** and **c** are 100 μm.

**Theoretical model and dynamic analysis of electro-capillary peeling.**

To clearly understand the electro-capillary peeling observed in the experiment and further explain the impact of the applied voltage and film's properties on the detaching behavior, we establish a theoretical model from the perspective of force analysis. In this work, a film with a thickness of $d_0$ is bonded on the ITO glass surface. At the pre-fabricated hole on the film, a liquid droplet with the density of $\rho$, surface tension of $\gamma$, relative permittivity of $\varepsilon_f$, and ion density of $\rho_f$ is placed (Fig. 4a). Because the working characteristic length ($10^{-4}$ m) in this work is much smaller than that of the Bond number ($10^{-3}$ m)[34], the effect of fluid gravity can be neglected. For the axisymmetric electro-capillary peeling mode, a unit model with a width of $l_0$ is analyzed in polar coordinates. $r(t)$ and $w(r)$ are the peeling length and the deflection of film, respectively. Calculations based on the theory of elasticity[38], the deflection of film is described as

$$T\frac{d^2 w}{dr^2} = p, \tag{1}$$

where $p$ is the net pressure loading on the film and $T$ is the film tension stress. At the strain of $\varepsilon$, $T = E\varepsilon d_0$, where $E$ is the elastic modulus of film. Under electric fields, the net pressure loading on the film is obtained from the equilibrium relationship between the film's dead weight and the electric force, which is expressed as $p = T^M - \rho_m g d_0$, where $T^M$, $g$, and $\rho_m$ are the Maxwell stress, gravitational acceleration, and film's density, respectively. According to the Korteweg-Helmholtz law[39], the Maxwell stress is expressed as $T^M = \varepsilon_f \rho_f E_e$, where $E_e$ is the electric intensity. Due to the film with a micron thickness, the pressure induced by the film's dead weight can be neglected compared with Maxwell stress. Substituting the net pressure into Eq. (1), we have

$$T\frac{d^2w}{dr^2} = \varepsilon_f \rho_f E_e. \tag{2}$$

In this work, the electric intensity is expressed as $E_e = U/\sqrt{r^2 + z_1^2}$, where $z_1$ is the distance from the positive electrode to the glass surface and $U$ is the applied voltage. Combining Eq. (2) and the boundary condition of $w|_{r=r_0} = (dw/dr)|_{r=r_0} = 0$, the deflection of film in the electro-capillary peeling

$$w(r) = k_1 r \left( \ln \frac{r}{r_0} - 1 \right) + k_1 r_0, \tag{3}$$

where $k_1 = \varepsilon_f \rho_f U / T$, $r_0$ is the peeling length at present. Due to the applied voltage being ultra-small and the electrode being far away from the film, we consider that the voltage $U$ on the film would degenerate into the zeta potential $\zeta$ with evolution, and $k_1$ would become that $k_2 = \varepsilon_f \rho_f \zeta / T$. Therefore, the lifting height does not vary obviously when the voltage changes from 2.5 to 4.5 V (Corresponding to Fig. 2d).

Considering that the peeling process of film on the rigid substrate is similar to the fracture. In this process, Maxwell stress competes with the film's tension stress and adhesive stress to realize the film's detachment. The adhesive energy of bonding layer is expressed as $U_a = f_{adhesion} r_0 l_0$, where $f_{adhesion}$ is the adhesive stress between the film and substrate. In addition, it is assumed that a state of plane strain exists on the film and the elastic energy is described as $U_e = E\varepsilon r_0 l_0 d_0$. At the onset of electro-capillary peeling, the liquid cannot wet the bonding layer due to the large Laplace pressure (Fig. 4b). Under such conditions, the action length of electric force is the width of induced charge aggregation. Therefore, the work of Maxwell stress is described by $U_p = F^M z_0 = T^M l_0 \lambda_D k_1 r_0$, where $F^M$ is the electric force and $\lambda_D$ is the Debye length. According to the principle of energy minimization, we can obtain the critical electric force that

$$F^M = \frac{1}{k_2} \left( E\varepsilon l_0 d_0 + f_{adhesion} l_0 \right). \tag{4}$$

After neglecting the term of $O(f_{adhesion}l_0)$ and substituting $k_2 = \varepsilon_f \rho_f \zeta / T$ into Eq. (4), the critical peeling voltage

$$U_c = \frac{E \varepsilon d_0 r_0 f_{adhesion}}{\lambda_D \varepsilon_f^2 \rho_f^2 \zeta}. \tag{5}$$

We infer from Eq. (5) that the critical peeling voltage is of order with the thickness and elastic modulus ($U_c \propto d_0$, $U_c \propto E$). The critical peeling voltage would be greater for the harder or thicker film. Observations shown in Fig. 4c further demonstrate this speculation that the critical peeling voltage of electro-capillary peeling is proportional to the film's thickness and elastic modulus, which is consistent with Eq. (5).

Substituting Eq. (3) and the net pressure into Poiseuille law, we obtain the liquid droplet spreading rate

$$w^2 \frac{dp}{dr} = \mu v, \tag{6}$$

where $\mu$ and $v$ are the viscosity and spreading rate of liquid droplet, respectively. It has demonstrated that the relationship between the film deflection and the spreading length can be described by $w(r) \sim r$. To find the spreading law, we substitute the net pressure into Eq. (6), and the resulting relation is that

$$r \sim \frac{\varepsilon_f^3 \rho_f^3 \zeta^2 U t}{\mu E^2 d_0^2}, \tag{7}$$

where $t$ is the peeling time. The resulting relation indicates that the peeling length versus time is $r \sim t$, and the impact of applied voltage, the thickness and elastic modulus on the peeling length are described by $r \sim U$, $r \sim d_0^{-1/2}$ and $r \sim E^{-1/2}$, respectively. Experimental observations of peeling length versus time, applied voltage, and film's properties are shown in Figs. 4d, e. The solid line is the slope proposed in theory and discrete points are the peeling length observed in the experiment. The peeling length is proportional to the peeling time, and the impact of applied voltage and film's properties on the electro-capillary peeling are all consistent with the theoretical analysis. Within the uncertainties, an excellent agreement is shown between the predicted and experimental data.

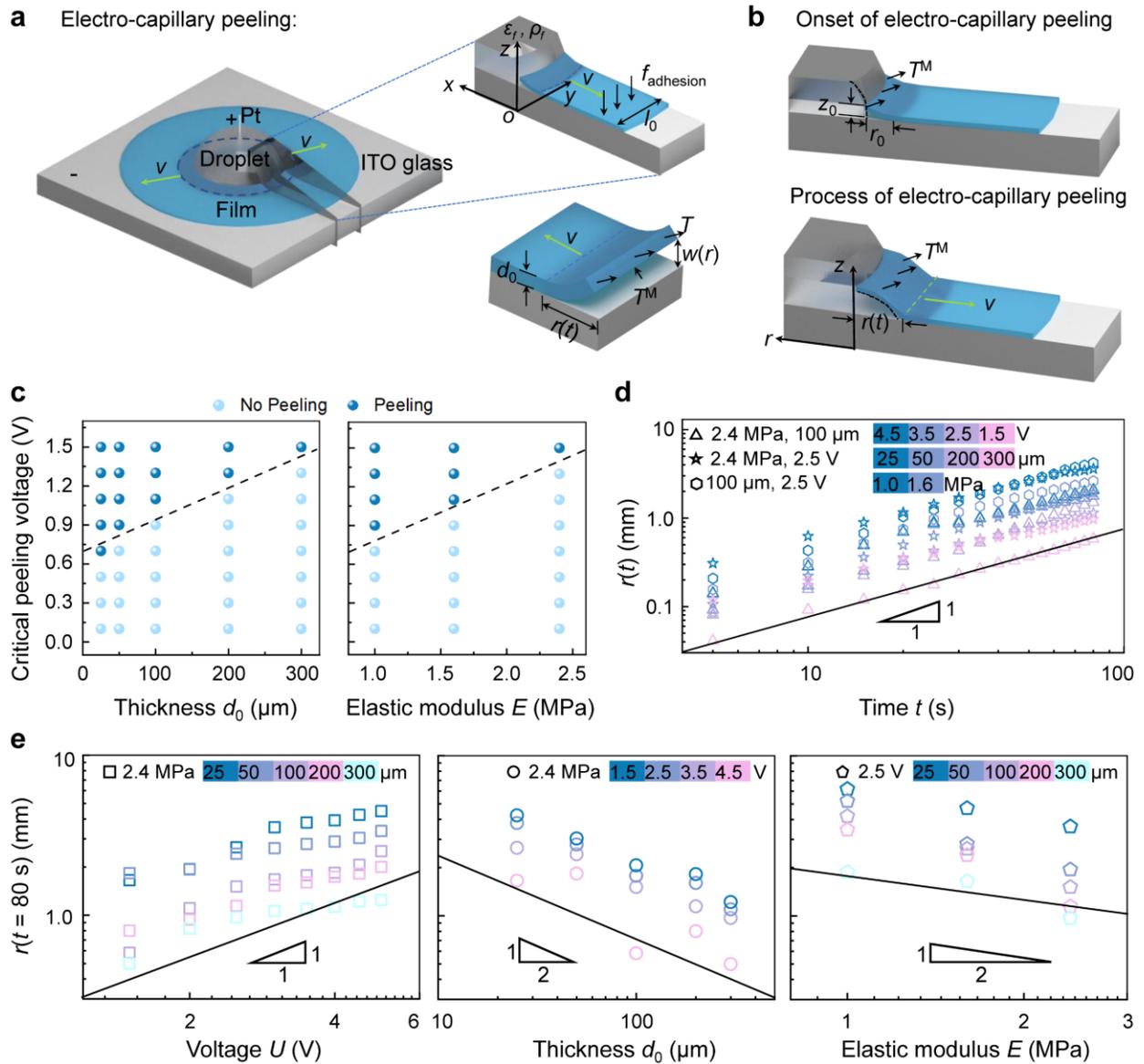

**Fig. 4| Working mechanism of electro-capillary peeling. a** Schematic illustration of electro-capillary peeling mechanism. A unit model with a width of $l_0$ is analyzed in orthogonal coordinates. $x$, $y$, and $z$ are the radial, tangential, and film thickness directions, respectively. **b** Electro-capillary peeling process. At the onset of electro-capillary peeling, the liquid droplet detaches the film with an ultra-low lifting height of $z_0$ and does not wet the bonding layer. As the peeling proceeds, the lifting height increases, and the liquid droplet would wet and spread into the bonding layer. The droplet outline is marked by a dotted line. **c** Critical peeling voltage of electro-

capillary peeling for films with various thicknesses and elastic moduli. Dark and light blue spheres represent that films can be peeled and not be peeled at such voltage. **d** Peeling law of the electro-capillary peeling versus time. Solid line: $r(t) \sim t$. **e** The impact of applied voltage and film's properties on the peeling length. Solid line: $r(t) \sim U$, $r(t) \sim d_0^{-1/2}$, and $r(t) \sim E^{-1/2}$. All experiments were repeated at least 6 times.

**Electro-capillary peeling with ultra-low strain for functional devices' protection.**

Due to the long-term reliability being a major challenge for the commercialization of functional devices on the film, developing a nondestructive method for peeling and transferring is significant in these applications. Unlike the traditional approach, the electro-capillary peeling method has a novel detaching mode that lifts films from the substrate by taking advantage of driving liquid to percolate and spread into the bonding layer. Characterized by using the three-dimensional digital image correlation (3D DIC, see Methods), the displacement and strain fields are presented in Figs. 5a, b (Supplementary Movie 4). The film's displacement is only 0.152 mm in the $z$-direction and its strain is less than 0.00332 during the peeling process, which are both ultra-low. Compared with the traditional method, the film detached by the electro-capillary peeling method has a much lower strain than that peeled by the debonded strip and blister method (Fig. 5c). At the same peeling rate, the strain of film peeled by the blister approach is almost six times as much as that detached by the electro-capillary peeling method. Figure 5d shows the performance of ZnO nano-rods on the film after detaching by using the electro-capillary peeling method and debonded strip. Observations present that cracks occurred on the ZnO nano-rods layer when the film was peeled by using the deboned strip ten times. Yet, the ZnO nano-rods layer was adequately preserved in the electro-capillary peeling method. In addition, previous studies have proven that many micro-/nano-devices on the film would be ruined due to the large strain[40, 41]. For this reason, the electro-capillary peeling method would be a valid approach to protect the functional device in these applications.

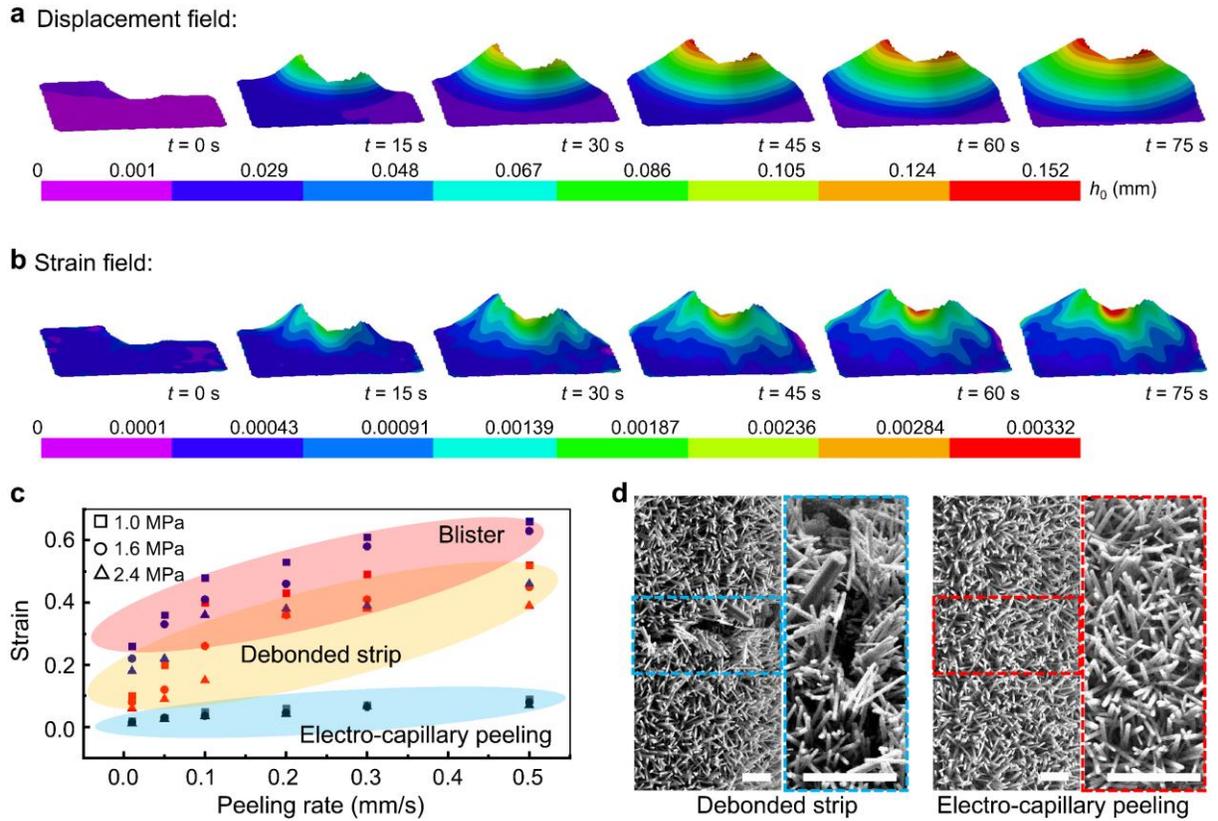

**Fig. 5| Deformation of film during the electro-capillary peeling process. a-b** Displacement and strain fields of film detached using the electro-capillary peeling method. Displacement and strain fields exhibit deformation in the $z$-direction. Color bars represent the displacement and strain values. **c** Strain of film versus peeling rate in various methods. Debonded strip method peels the film with an angle of 90°. The thickness of tested films is 100 μm. **d** Characterization of ZnO nano-rods on the film after peeling ten times. All scale bars are 100 μm.

**Discussion**

Above, we have provided experimental evidence of the electro-capillary peeling method that the liquid is driven to percolate and spread into the bonding layer under electric fields for the film's detachment. The electro-capillary peeling method first demonstrates that the electric field controls the liquid to peel off films from the substrate.

Evaluated by various applied voltages, observation indicated that the detaching rate of electro-capillary peeling method could be actively controlled by the applied electric field (Corresponding to Fig. 2). Depending on the application, the peeling rate and detaching length would be precisely altered by the applied voltage. Systematic experiments conducted with different films further reveal that, although the performance of the electro-capillary peeling method is affected by the film's thickness, elastic modulus, and type, this approach is clearly appropriate in a broad range of films (Corresponding to Fig. 3). Additionally, the acquired deformation field highlights that the film has an ultra-low strain when it is detached by using the electro-capillary peeling method. For the long-term reliability of functional devices on thin films, the electro-capillary peeling method would be a wise option for detaching and transferring in applications (Corresponding to Fig. 5). Ultimately, the experimental setup and critical peeling voltage indicate that the film can be easily detached by using electro-capillary peeling method with a simple apparatus at only 0.7 V (Corresponding to Figs. 1 and 4). This result would also denote that films would be detached at the voltage of an AAA battery in practice, which is quite appealing.

Although many approaches have been proposed to detach thin films from the substrate[22-26], the electro-capillary peeling method is fundamentally different from them. Firstly, the electro-capillary peeling method is a simple physical detaching approach. Unlike the modification of bonding layer, the electro-capillary peeling method does not change the interfacial properties of thin films and does not need complicated chemical preparation. It extremely protects the film's features in applications. Furthermore, the electro-capillary peeling method is an active control technique and is appropriate in a broad range of films. Compared with the capillary peeling[29, 30], the electro-capillary peeling method can be actively altered by the applied voltage and applied in hydrophobic, hydrophilic, and even superhydrophilic films. Corresponding to Fig. 3 and Supplementary Table S1, the electro-capillary peeling method readily detaches the film with adhesive stress from 2.1 to 32.3 N/m.

Besides, the electro-capillary peeling method can be used in many solutions. Extended experimental observations performed in $CaCl_2$, $CuSO_4$, LiCl, NaCl, KOH, and NaOH solutions with a concentration of 0.1, 0.3, 0.5, and 1.0 mol/L demonstrate that the electro-capillary peeling method can be applied in neutral, acidic, and alkaline solutions of wide-range concentrations (Supplementary Figs. S6, S7). Last but not least, the electro-capillary peeling method also easily detaches fully attached micro/nanofilms due to the special peeling mode. With the wide use of micro/nanofilms, the electro-capillary peeling would be an appealing approach for the film's detaching, transferring, and reusing in applications.

**Conclusions**

Herein, we propose an electro-capillary peeling method for the thin film's detachment. Unlike the traditional detaching approach, the electro-capillary peeling method achieves the film's detachment by driving liquid to percolate and spread into the bonding layer under electric fields. This is the first demonstration that electric field controls the liquid to detach films from substrates. Evaluated by various applied voltages and films, it is demonstrated that the electro-capillary peeling method is an active control technique and is appropriate in a broad range of films. Analyzed from the perspective of loading force, it is indicated that the electro-capillary peeling method is actualized by utilizing Maxwell stress under electric fields competing with the film's adhesive stress and tension stress. The peeling length versus time, applied voltage, film's thickness, and elastic modulus are described by $r \sim t$, $r \sim U$, $r \sim d_0^{-1/2}$, and $r \sim E^{-1/2}$, respectively. Furthermore, the film's deformation shows that the film detached by the electro-capillary peeling method with a much lower strain for functional devices' protection. Critical peeling voltage reveals that the film can be readily detached at only 0.7 V in many solutions of various concentrations. For the advantages, this work exemplifies the new method of "electro-capillary peeling" that would be attracted much interest from academics and industry and might promote "water-based peeling" applied in practical applications.

**Methods**

**Characterization of electro-capillary peeling for thin films.** Thin film is a key base material for many applications, such as flexible electronics[1, 2], soft robotics[3, 4], micro/nano devices[5, 6], and so on. Although some actuation approaches[21-26, 29] have been proposed to peel films from substrates, these methods always require restricted conditions to limit the widespread use of films. Therefore, developing a simple and non-destructive peeling method for films is a critical technology in these applications. In the electro-capillary peeling method, the experimental setup is schematically shown in Fig. 1. Before each test, ITO glass (resistivity < 6 Ω, and 10 cm in diameter) was rinsed in alcohol solution (99.97%) to remove the dust. After the PDMS film was treated by using the Plasma method, it was fabricated a hole (2 mm in diameter) at the center and adhered to the ITO glass surface. A micro pump (Pump 11 Elite, Harvard Apparatus, USA) injected an electrolyte droplet (1.0 mol/L KCl solution) with a volume of 20 μL into this pre-fabricated hole, and then an electric field would be applied to the droplet by a DC supply power (LPS3020D, Lodestar, China). One of the Pt wire electrodes (99.99%, 200 μm in diameter) was inserted vertically into the liquid droplet and connected to the positive pole, and the other Pt wire electrode was pasted on the ITO glass surface and attached to the negative pole. Observation of electro-capillary peeling was recorded by two cameras (JHUM1204s-E, Jinghang, China) mounted on an independent *XYZ* stage at 30 frames per second. The displacement of the droplet was obtained by using Image J software.

**Preparation of PDMS film.** A kit including PDMS base and agent (Sylgard 184, Dow corning) was used to prepare PDMS films. The liquid PDMS mixture, consisting of the base and curing agent with a weight ratio of

10:1, was degassed and poured into a petri dish. A fixed volume of liquid PDMS mixture (8 mL) was deposited onto the center of the silicon wafer followed by a spin coating step. The PDMS film had a uniform thickness immediately after spin-coating with 30 s by a spin coater (WS-400B-6NPP-LITE, Laurell, USA). In this step, the thickness of PDMS was obtained with a designed rate. PDMS films with a thickness of 25, 50, 100, 200, and 300 μm were fabricated with spin rates at 3100.0, 1250.0, 650.0, 420.0, and 270.0 rpm/s, respectively. This PDMS film is cured on a leveled hot plate and finally transferred to a box furnace. After curing for 100 minutes at 65 °C, the PDMS is ready for further experiments. Additionally, PDMS films with an elastic modulus of 1.0, 1.6, and 2.4 MPa are achieved by consisting of the base and curing agent with a weight ratio of 20:1, 15:1, and 10:1, respectively. Other operations are the same as above.

**Fabrication of ZnO nano-rods on PDMS films.** To explore the electro-capillary peeling method in the protection of nano-materials, ZnO nanorods were fabricated on the PDMS film. The crystal seed solution was prepared as follows: 5 g of $Zn(Ac)_2·2H_2O$, 0.8 g of monoethanol-amine (purchased from Shanghai Zhenpin Chemical Co., Ltd., China), and 20 mL of ethylene glycol monomethyl ether (purchased from Zhongshan Xinxin Chemical Co., Ltd., China) were mixed and stirred with a magnetic stirrer until the liquid became transparent. The growth liquid was achieved as follows: 0.4 g of hexamethylenetetramine (purchased from Beijing Lanyi Chemical Co., Ltd., China) and 0.82 g of $Zn(NO_3)_2·6H_2O$ (purchased from Beijing Lanyi Chemical Co., Ltd., China) were mixed in 100 mL deionized water and stirred to become transparent. The PDMS surface was covered by the crystal seed by dip-coating and treated at 320 °C for 2 minutes. The ZnO nanoseed was achieved on the PDMS surface. The PDMS surface with the growth liquid was placed into the reaction kettle (100 mL) at 95 °C for 12 h. The ZnO nanorods were planted on the film (Corresponding to Fig. 5d) and ready for further experiments.

**Preparation of electrolyte solution.** In the electro-capillary peeling method, we choose the KCl solution as the electrolyte solution, which had good electrical properties under electric fields. The KCl solution with a concentration of 0.1, 0.3, 0.5, and 1.0 mol/L was prepared by dissolving their powders (Sigma Aldrich Shanghai Trading Co., Ltd., China) in deionized water at room temperature. In addition, to explore the performance of electro-capillary peeling in various electrolyte solutions, we further fabricated $CaCl_2$, $CuSO_4$, LiCl, NaCl, KOH, and NaOH solutions with a concentration of 0.1, 0.3, 0.5, and 1.0 mol/L. The pH value of these electrolyte solutions was measured with a pH meter (PHSJ-4, INESA Scientific Instrument Co., Ltd., China) and the average value of six measurements was taken, which is shown in Supplementary Fig. 7.

**Electron microscopy and wettability characterization.** The topographies of the bonding layer, ZnO nano-rods, and film on the ITO glass surface were observed by environmental scanning electron microscopy (ESEM, Quanta FEG-250, FEI, USA) at a voltage of 10-15 kV. The contact angle of a liquid droplet with a volume of 10 μL on the film and ITO glass surface was tested by a video-based contact angle measuring device (DSA100, KRÜSS Scientific, Germany) with a precision of $\pm 0.1°$.

**Mechanical test of the film.** 1) Peeling test for films on the ITO glass surface. For a peel angle $\theta$, the adhesive force $F$ and the width of the film $l_0$, the adhesive stress $f_{adhesion}$ could be calculated by $f_{adhesion} = F(1-\cos\theta)/l_0$. In the case of 90° peel, $f_{adhesion}$ is equal to the peeling force $F$ divided by the width of the film $l_0$. We performed peeling experiments on the tensile test apparatus equipped with a 90° peeling device and a 10 N load cell. The peeling rate was controlled with a stepper motor with a precision of 0.001 mm/s. 2) Measurement of Young's modulus. Rectangular strips were cut from films (PDMS, Hydrogel, PET, and PEN) with dimensions of $80\,\text{mm} \times 20\,\text{mm} \times 1\,\text{mm}$. These rectangular strips were performed on a tensile test apparatus (All Around, Zwick Roell, Germany) equipped with a 100 N load cell. Samples were stretched at room temperature (20 °C) with no noticeable residual deformation and negligible hysteresis.

**Displacement and strain field of film.** Three-dimensional digital image correlation (3D DIC) with a spatial resolution of 10 μm was used to characterize the film's displacement and strain field in the electro-capillary peeling method (Supplementary Fig. S8). The micron particle was applied to sign the positional relation of the film in the 3D DIC. Observed by two high-speed cameras, the acquired image was analyzed in VIC-3D 9 system to characterize the film's deformation (Supplementary Fig. S8 and Supplementary Movie 4). The corresponding displacement and strain field are presented in Figs. 5a and b.

## Acknowledgments

This work was jointly supported by the National Natural Science Foundation of China (NSFC, Grant Nos. 12241205, 12032019, 11872363) and the National Key Research and Development Program of China.

## Data availability

The data supporting the findings of this study are available within the paper and its Supplementary files and are available from the corresponding author upon request.

## Author contributions

Y.-P. Z. conceptualized the idea and supervised the research. P. L. and X. H. conceived the idea and designed the experiment. And P. L. collected the datasets and drafted the manuscript. All authors read, contributed to the discussion, and approved the final manuscript.

## Competing interests

The authors declare no competing interests.

## Additional information

**Correspondence** and requests for materials should be addressed to Y.-P. Z.